\documentclass[runningheads, a4paper]{llncs}
\usepackage[T1]{fontenc}
\usepackage{graphicx}
\usepackage{listings}
\usepackage{amsmath}

\usepackage{multirow}
\usepackage[table,xcdraw]{xcolor}
\usepackage{tabularx}
\usepackage{booktabs}
\begin{document}

\title{Fault Injection in On-Chip Interconnects: A Comparative Study of Wishbone, AXI-Lite, and AXI}
%
%
\author{Hongwei Zhao, Vianney Lap\^otre, and Guy Gogniat}
\institute{Univ. Bretagne Sud, UMR 6285, Lab-STICC, Lorient, France}
\maketitle              
\begin{abstract}
Fault injection attacks exploit physical disturbances to compromise the functionality and security of integrated circuits. As System-on-Chip (SoC) architectures grow in complexity, the vulnerability of on-chip communication fabrics has become increasingly prominent. Buses, serving as interconnects among various IP cores, represent potential vectors for fault-based exploitation. In this study, we perform simulation-driven fault injection across three mainstream bus protocols—Wishbone, AXI-Lite, and AXI. We systematically examine fault success rates, spatial vulnerability distributions, and timing dependencies to characterize how faults interact with bus-level transactions. The results uncover consistent behavioral patterns across protocols, offering practical insights for both attack modeling and the development of resilient SoC designs.


\end{abstract}

\section{Introduction}
Embedded systems underpin a vast array of modern applications, ranging from consumer devices and industrial automation to automotive electronics and critical infrastructure. Their appeal lies in providing dedicated functionality with high efficiency, compactness, and cost-effectiveness. As their deployment expands into safety-critical and real-time domains, ensuring their security and dependability becomes paramount. Faults—originating from hardware imperfections, environmental stressors, or intentional interference—pose serious risks, potentially resulting in system failures, data corruption, or mission-critical breakdowns~\cite{bar2006sorcerer}~\cite{karaklajic2013hardware}. Consequently, enhancing the robustness of embedded systems remains a central focus for both academia and industry.

Fault injection has been extensively demonstrated on processors. For instance, Plundervolt attack \cite{9152636} shows for the first time that legitimate voltage scaling interfaces can themselves be abused to inject faults, successfully compromising even Intel SGX enclaves that are designed with strong security guarantees. Complementarily, \cite{boulifa2025countermeasures} surveys existing countermeasures deployed in embedded processors to mitigate such threats and improve system resilience. Another representative work, \cite{7774479}, describes an ARM-specific attack strategy that manipulates the program counter (PC) register when externally controlled data is loaded, illustrating how fault injection can directly subvert instruction flow in embedded systems.

Similarly, fault injection is also highly relevant in the context of memory systems. The work in \cite{8714811} analyzes attacker capabilities under physical access, discussing memory probing, prevalent attack models, and possible defenses. In \cite{101145}, the authors show that faulty operations can lead to misaddressed memory reads or writes and propose a protection scheme that validates every access to resist address tampering. More recently, \cite{schilling2022fipac} introduces FIPAC, a software-based control-flow integrity (CFI) mechanism that operates at the granularity of basic blocks, effectively strengthening execution security against both software and hardware fault injection.

Despite these advancements, system buses—critical for data exchange between processing cores, memory, and I/O—remain underexplored in fault resilience literature. This gap is noteworthy, given that buses often serve as shared resources and are exposed to both internal and external disturbances.

For example,~\cite{mishra2024faults} demonstrates that electromagnetic pulse (EMP) injections aimed at the bus are capable of compromising ARM TrustZone security. Most previous studies, however, focus on individual bus protocols in isolation. In contrast, this work performs a comparative fault analysis across three commonly adopted protocols—Wishbone, AXI-Lite, and AXI—through simulation-based injection. By stressing these protocols under fault conditions, we uncover recurring weaknesses and distill design guidelines for enhancing the robustness of on-chip communication. Similarly, \cite{6623558} investigated the consequences of fault attacks on both instruction and data buses. Yet, unlike our study, their work neither compared different bus architectures nor employed a systematic vulnerability-model traversal to explore and evaluate program-level impacts.

Compared with the analysis in \cite{zhao2024communication}, which examined fault injection vulnerabilities in the Wishbone bus, the present work extends the scope to include three widely used on-chip bus protocols—Wishbone, AXI-Lite, and AXI—under fault injection scenarios. While the earlier study focused exclusively on Wishbone, the current investigation reveals that AXI-Lite and AXI also exhibit structural weaknesses exploitable by fault-based attacks. Using simulation-driven campaigns, it compares critical aspects such as the types of registers targeted, the nature of data corruption, and the success rates of different fault models. The observed behaviors are further abstracted to identify recurring patterns across protocols, leading to a set of design recommendations aimed at improving bus-level resilience in embedded systems.

The primary contributions of this paper are summarized as follows:
\begin{enumerate}
\item A detailed structural analysis of signal transmission mechanisms across Wishbone, AXI-Lite, and AXI buses;
\item Empirical evaluation of fault injection outcomes across the three buses using four representative fault models;
\item Insights into how factors such as register types and timing windows affect the success rate of bus-targeted attacks;
\item Practical guidelines for system designers to enhance the robustness of communication interconnects.
\end{enumerate}

The remainder of this paper is structured as follows: Section~\ref{Data tranfer on the bus} provides a technical overview of the bus architecture and illustrates data flow behaviors before and after fault injection. Section~\ref{Environment configuration} outlines the experimental environment, including the automated fault injection framework and VerifyPin benchmark. Section~\ref{Vunerabilities on 3 Buses111} presents a comparative vulnerability analysis across the three bus protocols under multiple fault models. Finally, Section~\ref{Conclusion} concludes the paper and outlines future research directions.

\section{Background}
\label{Data tranfer on the bus}

Before performing a fault injection campaign, we examined the data transmission protocols of the Wishbone~\cite{wishbone_b3}, AXI-Lite~\cite{amba_axi}, and AXI~\cite{amba_axilite} buses in order to identify the control signals that could be manipulated by an attacker. In all three protocols, the transfer process is fundamentally governed by two categories of control signals: access selection (e.g., device or byte selection) issued by the master, and response signals (e.g., completion acknowledgment or error indication) returned by the slave. 

In Wishbone, transaction completion is indicated by \texttt{ACK}, with optional \texttt{ERR} and \texttt{RTY} for error or retry. Byte selection uses \texttt{SEL}, while slave choice is handled by external address decoding.

In AXI4-Lite, response signaling is carried on the \texttt{BRESP} (write response) and \texttt{RRESP} (read response) channels, which are two-bit fields indicating outcomes such as normal access success (OKAY), slave error (SLVERR), or decode error (DECERR). The associated handshake pairs (\texttt{BVALID}/\texttt{BREADY} and \texttt{RVALID}/\texttt{RREADY}) ensure that both master and slave agree on the completion of a transfer and on the validity of the response. Byte-level selection during writes is controlled by \texttt{WSTRB}, allowing partial word updates, while slave selection is not an explicit signal but is instead realized through address decoding in the interconnect fabric.

In AXI4, the same response scheme applies, with \texttt{BRESP} and \texttt{RRESP} carrying error or success codes and handshake signals coordinating acceptance. Additional burst control signals define transfer granularity and sequencing: \texttt{WLAST} and \texttt{RLAST} delimit the final beat of a burst, while \texttt{AWLEN}, \texttt{AWSIZE}, and \texttt{AWBURST} (and their read equivalents) describe the length, size, and type of the burst. \texttt{WSTRB} continues to indicate valid byte lanes on each data beat. Slave selection again derives from address decoding within the interconnect, but AXI4 extends this with full burst addressing semantics, enabling efficient block transfers across multiple slaves.

AXI4-Lite and AXI4 signal errors via the two-bit \texttt{BRESP} and \texttt{RRESP} fields, using codes such as normal access success (OKAY), slave error (SLVERR), and decode error (DECERR), together with the \texttt{BVALID}/\texttt{BREADY} or \texttt{RVALID}/\texttt{RREADY} handshakes. In the considered implementation, error states are handled by asserting the corresponding response code and forcing \texttt{RDATA} to all zeros, ensuring explicit fault indication and preventing propagation of invalid data.  

\section{Environment Configuration}
\label{Environment configuration}

To perform our experiments, we construct three SoC architectures with the LiteX framework~\cite{litex}. The target system adopts the RISC-V architecture and integrates a VexRiscv processor core~\cite{VexRiscv}. Three independent SoC configurations are instantiated, each employing a different interconnect protocol—Wishbone, AXI-Lite, and AXI—as the primary communication bus. The memory contains a ROM for instruction storage, an SRAM for runtime data, a MAIN\_RAM region reserved for user-defined memory, and a CSR unit responsible for the UART and timer. All memory blocks are zero-initialized prior to simulation.

We adopted the VerifyPin benchmark version V0 ~\cite{dureuil2016fissc} to investigate and leverage the behavior of fault injection attacks. In this program (Listing \ref{codev0}), the user input PIN, \texttt{g\_userPin}, is preset to "0000", \texttt{g\_cardPin}, is initialized to "4321". The comparison between these two variables simulates a typical failed authentication attempt at line 10. As a result, the flag variable \texttt{g\_authenticated} is set to "0" at line 12. A fault injection attack is considered successful if it causes \texttt{g\_authenticated} to be incorrectly set to "1" at line 11, thereby bypassing the authentication logic.

\begin{lstlisting}[caption={C code of VerifyPin function in benchmark V0}, label={codev0}, basicstyle=\ttfamily\footnotesize, numbers=left, numberstyle=\tiny, stepnumber=1, numbersep=5pt]
BOOL byteArrayCompare(UBYTE* a1, UBYTE* a2, UBYTE size){
    for(int i = 0; i < size; i++) {
        if(a1[i] != a2[i]) return 0;
    return 1;
    }
}
BOOL verifyPIN() {
    g_authenticated = 0;
    if(g_ptc > 0) {
        if(byteArrayCompare(g_userPin, g_cardPin, PIN_SIZE) == 1) 
            g_authenticated = 1;
        else g_authenticated = 0;
    }
}
\end{lstlisting}

To conduct our fault injection campaign, we utilized Questasim 10.6e as the simulation backend. Given the extensive number of test iterations needed to thoroughly evaluate architectural resilience, we adopted FISSA~\cite{PLG-24-dsd}, a Python-based framework designed to automate fault injection campaigns targeting microarchitectural components. 

After configuring the simulation environment, we defined attack condition. In line with the benchmark setup, the attacker is assumed to have no knowledge of the correct PIN ("4321") and operates under the assumption that the input PIN ("0000") does not match it. Thus, any successful authentication must arise solely due to fault-induced corruption. For this experiment, we assume an attacker targeting only the control registers connected to the system bus.

Although this control principle is consistent across protocols, the physical implementation in our LiteX-generated SoC differs from that of a native bus. In our design, the AXI-Lite and AXI interfaces are bridged through an internal Wishbone bus via LiteX protocol conversion modules. Moreover, the error handling mechanism specifically involves forcibly setting the ACK signal to one and clearing the output data.

The internal registers of all three buses are generated by LiteX. We present all targeted registers  in Table \ref{inject reg}. For Wishbone, fault injection targeted completion, arbitration, and selection logic, including \texttt{ACK}, \texttt{SEL}, and internal register \texttt{done} (connected with \texttt{DAT}, \texttt{ACK}, \texttt{error}), and \texttt{grant} (connected with \texttt{DAT}, \texttt{ADR}, \texttt{CYC}, \texttt{STB}, \texttt{WE}, \texttt{CTI}). For AXI‑Lite, representative state registers (connected with \texttt{AWREADY}, \texttt{WREADY}, \texttt{BVALID}, \texttt{BRESP}, \texttt{ARREADY}, \texttt{RVALID}, \texttt{RRESP}, \texttt{RDATA}) were selected across functional blocks, such as protocol FSMs (\texttt{fsm0}, \texttt{fsm1}), memory and bridge interfaces (\texttt{axilitesram0}, \texttt{axilite2csr}, \texttt{wishbone2axilite}), selection driver (\texttt{slave\_sel\_reg} connected with \texttt{AWREADY}, \texttt{WREADY}, \texttt{BVALID}, \texttt{BRESP}, \texttt{ARREADY}, \texttt{RVALID}, \texttt{DAT}), status flags (\texttt{last\_was\_read}, connected with \texttt{AWADDR}, \texttt{AWREADY}, \texttt{WREADY}, \texttt{BVALID}, \texttt{BRESP}), interconnect management (\texttt{rr\_read\_grant} connected with \texttt{ARVALID}, \texttt{ARREADY}, \texttt{ARADDR}, \texttt{ARPROT}, \texttt{RVALID}, \texttt{RREADY}), and completion flags (\texttt{cmd\_done} connected to \texttt{AWVALID} \texttt{BREADY} \texttt{ARVALID} \texttt{RREADY}, \texttt{data\_done} connected with \texttt{WVALID} and \texttt{BREADY}). For AXI, similar targets were used, with additional burst‑related and pipeline control signals (\texttt{ax\_beat\_first} connected with\\\texttt{ARVALID}, \texttt{AWVALID}, \texttt{ax\_beat\_last}, \texttt{last\_ar\_aw\_n} connected with \texttt{AWLEN} \\ and \texttt{AWSIZE}, \texttt{pipe\_valid\_source} connected with \texttt{AWLEN}, \texttt{AWSIZE}) absent in AXI‑Lite.

\begin{table}
\centering
\caption{All the registers targeted by fault injection in the 3 buses}
\label{inject reg}
\resizebox{\textwidth}{!}{
\begin{tabular}{|l|l|}
\hline
Bus & Registers \\
Wishbone & ACK, SEL, done, grant \\
AXI-Lite & state, selection driver, last\_was\_read, rr\_read\_grant, completion flags \\
AXI & same with AXI-Lite, ax\_beat\_first, ax\_beat\_last, last\_ar\_aw\_n, pipe\_valid\_source \\
\hline
\end{tabular}}
\end{table}

Registers unrelated to bus control—such as those for I/O, LEDs, clock/reset, timers, or holding program data—were excluded from fault injection, as they either bypass control logic or are protected by error‑detection mechanisms \cite{9316521,kiaei2020custom}, making effective injection unlikely.

To comprehensively evaluate system robustness, we introduced multiple fault models of varying complexity, all operating under a single-cycle fault assumption. In addition to a basic one-bit-flip model, three advanced fault models defined in FISSA were used:

\begin{enumerate}
\item \textbf{Bit-flip:} One bit-flip in a register.
\item \textbf{Manipulate Register:} Allows bit-flips within a single register. Both the number and location of bit changes are unconstrained, simulating a broad spectrum of register corruption scenarios.
\item \textbf{2 Bit-Flips:} Enforces exactly two bit-flips, either within one register or distributed across two.
\item \textbf{Manipulate Two Registers:} Injects arbitrary bit-flips into two separate registers simultaneously, capturing more complex situations involving concurrent register corruption.
\end{enumerate}

These four fault models can attack up to 4 bits of data. This type of attack has been demonstrated in \cite{colombier2021multi} to be precisely controllable using lasers.

Each simulation outcome was categorized into one of four classes:

\begin{enumerate}
\item \textbf{Crash:} The simulation terminated abnormally due to timeout or triggered crash behavior.
\item \textbf{Success:} The authentication logic was bypassed, with \texttt{g\_authenticated} incorrectly set to “1”.
\item \textbf{Change:} Memory content was altered, though no authentication success or detection occurred.
\item \textbf{Silence:} The fault had no observable effect—no authentication, detection, or state change.
\end{enumerate}

\section{Experimental results}
\label{Vunerabilities on 3 Buses111}
We conducted fault injection campaigns on all three bus protocols and analyzed the resulting system behavior under each fault model. Table~\ref{Vunerabilities on 3 buses} summarizes the observed outcomes, presenting the number of occurrences associated with each fault model across the Wishbone, AXI-Lite, and AXI buses.

\begin{table}
\centering
\caption{Fault injection results for each bus and fault model}
\label{Vunerabilities on 3 buses}
\begin{tabularx}{\textwidth}{|l|l|X|X|X|X|}
\hline
Bus & Fault model & Crash & Success & Change & Silence\\
\hline
                            & Bit-flip          & 47                        & 37                          & 124                         & 2366         \\
                            & Manipulate Register   & 320                       & 43                          & 161                         & 6496                              \\
                            & 2 Bit-Flips        & 547                       & 272                         & 914                        & 11137                 \\
\multirow{-4}{*}{Wishbone}  & Manipulate Two Registers & 5178                      & 778                         & 2891                       & 63225      \\
\hline
                            & Bit-flip          & 282                       & 4                           & 1913                       & 11577                     \\
                            & Manipulate Register   & 391                       & 6                           & 2589                       & 29942                   \\
                            & 2 Bit-Flips        & 10725                     & 153                         & 69811                      & 194831       \\
\multirow{-4}{*}{AXI-Lite}  & Manipulate Two Registers & 36215                     & 549                         & 224347                     & 1236105             \\
\hline
                            & Bit-flip          & 158                       & 4                           & 4833                       & 34269        \\
                            & Manipulate Register   & 435                       & 6                           & 6723                       & 90996       \\
                            & 2 Bit-Flips        & 14760                     & 333                         & 428382                     & 1421565                   \\
\multirow{-4}{*}{AXI}       & Manipulate Two Registers & 104655                    & 1328                        & 1514455                    & 9762850  \\
\hline
\end{tabularx}
\end{table}

The number of executed simulations increases from Wishbone to AXI‑Lite to AXI, reflecting greater architectural complexity and more control/status registers. Under simple fault models, Wishbone shows more successful attacks, indicating that lower‑complexity interconnects are more vulnerable to basic injections. The following presents a precise analysis of each faulted register combination that resulted in success.

Table \ref{Percentage on Wishbone} shows that, for Wishbone, most successful attacks hit the \texttt{ACK} control register whatever the considered fault model. The \texttt{SEL} register is particularly sensitive to the \textit{Manipulate Register} fault model. 

Table \ref{Percentage on Lite} shows that, for AXI-Lite, most successful attacks hit the \texttt{state} control register whatever the considered fault model. 

Table \ref{Percentage on AXI} shows that, for AXI, most successful attacks hit the \texttt{state} control register whatever the considered fault model. The \texttt{completion flag} register creates less successful attacks. 

\begin{table}
\centering
\caption{Distribution of successful register combination attacks under different fault models on the Wishbone Bus}
\label{Percentage on Wishbone}

\begin{tabularx}{\textwidth}{|l|X|X|X|X|}
\hline
Fault model & \texttt{ACK} & \texttt{SEL} & \texttt{ACK} \& \texttt{grant} & \texttt{ACK} \& \texttt{SEL} \\ 
\hline
Bit-flip & 94.59\% & 5.41\% & - & - \\
Manipulate Register & 81.40\% & 18.60\% & - & - \\
2 Bit-Flips & 94.12\% & 3.68\% & 1.10\% & 1.10\% \\
Manipulate Two Registers & 85.73\% & 12.34\% & 0.38\% & 1.55\% \\
\hline
\end{tabularx}
\end{table}

\vspace{-1cm}
\begin{table}
\centering
\caption{Distribution of successful register combination attacks under different fault models on the AXI-Lite Bus}
\label{Percentage on Lite}

\begin{tabularx}{\textwidth}{|l|X|X|X|}
\hline
Fault model & state & selection driver \& state & completion flag \& state \\
\hline
Bit-flip & 100.00\% & - & - \\
Manipulate Register & 100.00\% & - & - \\
2 Bit-Flips & 98.02\% & 1.32\% & 0.66\% \\
Manipulate Two Registers & 98.36\% & 1.46\% & 0.18\% \\
\hline
\end{tabularx}

\end{table}

\vspace{-1cm}
\begin{table}
\centering
\caption{Distribution of successful register combination attacks under different fault models on the AXI Bus}
\label{Percentage on AXI}
\begin{tabularx}{\textwidth}{|l|X|X|X|}
\hline
Fault model & state & completion flag & completion flag \& state \\
\hline
Bit-flip & 75.00\% & 25\% & - \\
Manipulate Register & 83.33\% & 16.67\% & - \\
2 Bit-Flips & 60.00\% & 20\% & 20.00\% \\
Manipulate Two Registers & 83.13\% & 16.79\% & 0.08\% \\
\hline
\end{tabularx}
\end{table}

\vspace{-0.5cm}
\subsection{Fault Characterization and Mitigation Approaches}
Based on our experimental observations, we categorized the possible fault-induced behaviors for the observed successful attacks across the three bus types into the following classes:

\begin{itemize}
\item \textbf{Instruction skip:} The CPU omits the execution of a scheduled instruction or data. For example, a fault in the \texttt{ACK} register may disrupt the timing of data or instruction fetching, causing the CPU to miss the comparison instruction. As a result, the key verification step on line~10 of Listing~\ref{codev0} can be skipped.  

\item \textbf{Data reset:} The targeted instruction or data is overwritten with its default value (typically zero). For instance, a fault in the \texttt{state} register may force the state machine into an error-handling mode, which resets the bus output. Consequently, the CPU reads an all-zero value for the variable \texttt{g\_cardPin}, making it identical to \texttt{g\_userPin}.  

\item \textbf{Data misread:} The CPU retrieves data from an unintended address. For example, a fault in the \texttt{SEL} signal can cause the CPU to fetch data from \texttt{CSR} instead of \texttt{SRAM}. If the value retrieved for the variable \texttt{g\_userPin} is replaced by \texttt{g\_cardPin}, the authentication process can be bypassed.  

\item \textbf{Data multiread:} The CPU simultaneously fetches data from multiple address locations, producing inconsistent or unintended results. For example, a fault in the \texttt{SEL} signal may cause the CPU to read from both \texttt{CSR} and \texttt{SRAM}, with the result computed as the logical OR of the two values. This behavior may cause the value of \texttt{g\_userPin} to be corrupted and interpreted as \texttt{g\_cardPin}.  
\end{itemize}

Based on the previous Tables \ref{Percentage on Wishbone}, \ref{Percentage on Lite}, and \ref{Percentage on AXI}, our analysis shows that register attacks can succeed through multiple register types.

For the Wishbone bus, the root cause of these vulnerabilities lies in its architectural design. The \texttt{ACK} signal directly controls data transfers, yet no timing violation detection is implemented at the bus level. In contrast, the \texttt{SEL} signal selects among four memory blocks via combinational logic, meaning that if compromised, it can arbitrarily redirect data access. Empirical results indicate that \texttt{ACK}-related attacks are more prevalent than those targeting \texttt{SEL}. The \texttt{ACK} signal is register-based and represented by a 4-bit value; flipping any bit can disrupt data transfer, thereby enlarging the attack surface and increasing the probability of success. By comparison, \texttt{SEL}-based attacks are more constrained, with their effectiveness heavily dependent on memory initialization states, memory layout strategies, and the behavior defined by the underlying assembly code.

On the AXI-lite bus, the \texttt{state}, \texttt{selection driver}, and \texttt{completion} registers collectively govern the data transfer mechanism, enabling potential attack vectors. The \texttt{state} register controls state machine transitions, thereby influencing control signal data and addresses. The \texttt{selection driver} affects the selection signal, indirectly determining which memory module the CPU accesses. The \texttt{completion} register can influence updates to the \texttt{state} register.

Attacks targeting the \texttt{state} register exhibit significantly higher success rates than those aimed at the other two registers. This is primarily because the AXI-lite bus contains a large number of \texttt{state} registers—far exceeding the single \texttt{selection driver} and two \texttt{completion} registers combined. Moreover, the \texttt{state} register can directly force the state machine into error-handling mode, compelling the CPU to read all-zero data. This reset can convert card pins into user pins, thereby contributing to the high success rate.

The AXI bus demonstrates similar characteristics as AXI-lite. The abundance of \texttt{state} registers and robust error-handling mechanisms result in consistently high success rates when targeting the \texttt{state} registers. \texttt{Completion} registers achieve moderate success by indirectly influencing \texttt{state} registers. However, due to stricter data control in AXI, compromising the \texttt{SEL} signal requires simultaneous manipulation of the \texttt{selection driver} and two additional control registers—a condition our fault model cannot meet—resulting in zero successful attacks.

Table~\ref{Percentage data} presents the proportion of attack effect on instruction and data vulnerabilities observed under the \textit{Manipulate Two Registers} fault model across the three evaluated bus protocols.

\begin{table}
\centering
\caption{Percentage of Manipulate Two Registers fault attacks that modify the data or instruction read operation across the three bus architectures}
\label{Percentage data}

\begin{tabularx}{\textwidth}{|l|X|X|X|}
\hline
& Wishbone & AXI-Lite & AXI \\
\hline
data related & 17.22\% & 97.27\% & 100\% \\
instruction related & 82.78\% & 2.73\% & 0\% \\
\hline
\end{tabularx}

\end{table}

\vspace{-0.5cm}
On the Wishbone bus, a high ratio of faults impacts instruction execution. This is primarily due to frequent corruption of instructions such as \texttt{lui} and \texttt{mv}, which play key roles in data movement and initialization. As these instructions are frequently targeted, the instruction-related fault ratio is notably elevated.

During our experiments, frequent occurrences of data reset were observed on the AXI-Lite and AXI buses, revealing a distinct fault pattern. On the Wishbone bus, when the control signal \texttt{done} is enabled, the bus data are initialized to all “1”. Since this value does not correspond to any valid variable or instruction, such resets cannot directly satisfy a success condition. By contrast, in AXI-Lite and AXI systems generated by LiteX, two mechanisms may reset the bus data to zero: lightweight error-handling routines and the behavior of state signals. In our benchmark scenario, the user PIN is set to “0000”. As a result, a fault-induced reset can lead to interpreting the stored card PIN as “0000” instead of "4321" due to the reset, thereby enabling unauthorized authentication. In these settings, data corruption—rather than instruction corruption—emerges as the dominant attack pathway.

Additionally, across all three bus protocols, we observed that fault effects for all four injection models consistently occurred during CPU data read phases. This pattern is attributed to the structure of the \texttt{VerifyPin} benchmark, which includes very few write operations. Moreover, the registers responsible for handling write transactions on each bus are limited in number and infrequently accessed, making them less exposed to injection and thus reducing the likelihood of write-related fault effects.

The proposed classifications serve as a basis for identifying common fault manifestations (i.e., instruction skip, data reset, data misread and data multiread) and evaluating their severity across different bus protocols. 
Table \ref{compare bus} provides the list of the fault effects for each fault model and for the three buses. Note that the above models can be interpreted differently under certain conditions. For example, if the data reset has no effect, it can be, in certain case, regarded as an instruction skip. Another example is, if the data read operation, due to the fault, leads to 0, then the data multiread model can be treated as a data reset model.

\begin{table}
\centering
\caption{Possible fault exploitation for the three different buses depending on the fault model}
\label{compare bus}

\begin{tabularx}{\textwidth}{|l|X|X|X|}
\hline
                                          & Wishbone         & AXI-Lite         & AXI          \\
\hline                                          
\multirow{3}{*}{Bit-flip}                 & instruction skip & data reset       & data reset   \\
                                          & data reset       &                  & data misread \\
                                          & data multiread   &                  &              \\
\hline                                          
\multirow{4}{*}{Manipulate Register}      & instruction skip & data reset       & data reset   \\
                                          & data reset       & data misread     & data misread \\
                                          & data misread     &                  &              \\
                                          & data multiread   &                  &              \\
\hline                                          
\multirow{4}{*}{2 Bit-Flips}              & instruction skip & instruction skip & data reset   \\
                                          & data reset       & data reset       & data misread \\
                                          & data misread     & data misread     &              \\
                                          & data multiread   & data multiread   &              \\
\hline                                          
\multirow{4}{*}{Manipulate Two Registers} & instruction skip & data reset       & data reset   \\
                                          & data reset       & instruction skip & data misread \\
                                          & data misread     & data misread     &              \\
                                          & data multiread   & data multiread   &             \\
\hline                                          
\end{tabularx}

\end{table}

\vspace{-1cm}
\subsection{Design recommendations for more resilient systems}

Based on the observed vulnerabilities and fault injection results, we propose the following recommendations for circuit and system designers:

\begin{itemize}
\item \textbf{Prioritize protection of control registers:} Particular attention should be given to safeguarding data path control elements such as handshake and state registers. Although complex bus protocols inherently resist simplistic fault models due to their layered behavior, they often include a larger number of registers—making them potentially more susceptible to sophisticated or exhaustive fault injection strategies.

\item \textbf{Reinforce handshake logic:} Faults targeting respond-related signals (e.g., \texttt{ACK}) tend to induce instruction skipping with relatively simple fault models, whereas faults on chip-select signals (e.g., \texttt{SEL}) typically result in incorrect memory access. Therefore enforced handshake logic warrants stronger protection mechanisms.

\item \textbf{Go beyond traditional error handling:} Conventional bus-level designs often emphasize error detection and recovery, but may overlook active fault injection scenarios. We recommend that fault tolerance strategies be re-evaluated under adversarial conditions, with particular emphasis on minimizing the propagation effect of fault-handling logic.

\item \textbf{Mitigate effects of zero-value initialization:} Abnormal inputs such as all-zero patterns should be identified and restricted during critical data flow stages. Sensitive values—such as passwords or authentication codes—should be guarded using masking techniques or write-protection features to prevent forced zeroing. Additionally, authentication processes should incorporate consistency checks to detect and flag suspicious bypass behavior.
\end{itemize}

Our experiments demonstrate that fault tolerance mechanisms on complex buses are insufficient to address fault injection, necessitating the implementation of countermeasures. To defend against the four fault models considered in this study, we recommend applying triple modular redundancy (TMR) to registers shown to be vulnerable in our experiments. Furthermore, instead of traditional combinational \texttt{SEL} logic, we advocate for using multiplexers between selection signals and memory blocks. This can effectively prevent unintended multiple memory accesses, thereby mitigating fault effects related to data multireads.

\section{Conclusion}
\label{Conclusion}
This work systematically compared Wishbone, AXI‑Lite, and AXI under multiple fault models, revealing that while complex buses better resist simple faults, their additional control/status registers create new attack surfaces. Analysis of affected registers, corruption types, and fault timing exposed shared and protocol‑specific weaknesses, leading to design recommendations such as protecting handshake signals, hardening against instruction‑skips, and refining selection logic. 

In future work, we aim to explore hybrid countermeasures that combine hardware and software defenses to mitigate the identified vulnerabilities with minimal resource overhead. In particular, we plan to evaluate the effectiveness of such techniques against more advanced fault models, including multi-cycle or temporally correlated fault injections targeting multiple execution phases.



\begin{thebibliography}{99}

\bibitem{bar2006sorcerer}
Bar-El, H., Choukri, H., Naccache, D., Tunstall, M., Whelan, C.: 
The sorcerer's apprentice guide to fault attacks. 
Proceedings of the IEEE \textbf{94}(2), 370--382 (2006)

\bibitem{karaklajic2013hardware}
Karaklaji{\'c}, D., Schmidt, J.-M., Verbauwhede, I.: 
Hardware designer's guide to fault attacks. 
IEEE Trans. VLSI Systems \textbf{21}(12), 2295--2306 (2013)

\bibitem{9152636}
Murdock, K., Oswald, D., Garcia, F.D., Van Bulck, J., Gruss, D., Piessens, F.: 
Plundervolt: Software-based fault injection attacks against Intel SGX. 
In: IEEE Symp. Security and Privacy (SP), pp. 1466--1482. IEEE (2020). 
\doi{10.1109/SP40000.2020.00057}

\bibitem{boulifa2025countermeasures}
Boulifa, R., Di Natale, G., Maistri, P.: 
Countermeasures against fault injection attacks in processors: A review. 
Information \textbf{16}(4) (2025)

\bibitem{7774479}
Timmers, N., Spruyt, A., Witteman, M.: 
Controlling PC on ARM using fault injection. 
In: Workshop on Fault Diagnosis and Tolerance in Cryptography (FDTC), pp. 25--35. IEEE (2016).
\doi{10.1109/FDTC.2016.18}

\bibitem{8714811}
Werner, M., Schilling, R., Unterluggauer, T., Mangard, S.: 
Protecting RISC-V processors against physical attacks. 
In: DATE Conference, pp. 1136--1141. IEEE (2019).
\doi{10.23919/DATE.2019.8714811}

\bibitem{101145}
Schilling, R., Werner, M., Nasahl, P., Mangard, S.: 
Pointing in the right direction – securing memory accesses in a faulty world. 
In: ACSAC 2018, pp. 595--604. ACM (2018). 
\doi{10.1145/3274694.3274728}

\bibitem{6623558}
Moro, N., Dehbaoui, A., Heydemann, K., Robisson, B., Encrenaz, E.: 
Electromagnetic fault injection: Towards a fault model on a 32-bit microcontroller. 
In: FDTC, pp. 77--88. IEEE (2013).
\doi{10.1109/FDTC.2013.9}

\bibitem{schilling2022fipac}
Schilling, R., Nasahl, P., Mangard, S.: 
FIPAC: Thwarting fault- and software-induced control-flow attacks with ARM pointer authentication. 
In: COSADE, pp. 100--124. Springer (2022)

\bibitem{mishra2024faults}
Mishra, N., Chakraborty, A., Mukhopadhyay, D.: 
Faults in our bus: Novel bus fault attack to break ARM TrustZone. 
In: NDSS Symposium (2024)

\bibitem{mitic2006survey}
Miti{\'c}, M., Stoj{\v{c}}ev, M.: 
A survey of three system-on-chip buses: Amba, coreconnect and wishbone. 
In: Proc. 41st Int. Conf. Inform. Commun. Energy Syst. Technol. (ICEST), pp. 282--285 (2006)

\bibitem{zhao2024communication}
Zhao, H., Lapotre, V., Gogniat, G.: 
Communication architecture under siege: An in-depth analysis of fault attack vulnerabilities and countermeasures. 
In: IEEE Int. Conf. Cyber Security and Resilience (CSR), pp. 890--896. IEEE (2024)

\bibitem{litex}
Kermarrec, F., Bourdeauducq, S., Le Lann, J.-C., Badier, H.: 
LiteX: an open-source SoC builder and library based on Migen Python DSL. 
arXiv preprint arXiv:2005.02506 (2020)

\bibitem{VexRiscv}
Papon, C.: 
VexRiscv: A modular RISC-V CPU generator. 
\url{https://github.com/SpinalHDL/VexRiscv}, accessed: 2025-04-01

\bibitem{dureuil2016fissc}
Dureuil, L., Petiot, G., Potet, M.-L., Le, T.-H., Crohen, A., de Choudens, P.: 
FISSC: A fault injection and simulation secure collection. 
In: SAFECOMP, pp. 3--11. Springer (2016)

\bibitem{PLG-24-dsd}
Pensec, W., Lap{\^o}tre, V., Gogniat, G.: 
Scripting the unpredictable: Automate fault injection in RTL simulation for vulnerability assessment. 
In: Euromicro DSD, pp. 369--376. IEEE (2024)

\bibitem{9316521}
Sakamoto, J., Hayashi, S., Fujimoto, D., Matsumoto, T.: 
How to code data integrity verification secure against single-spot-laser-induced instruction manipulation attacks. 
In: AICCSA, pp. 1--8. IEEE (2020).
\doi{10.1109/AICCSA50499.2020.9316521}

\bibitem{kiaei2020custom}
Kiaei, P., Mercadier, D., Dagand, P.-E., Heydemann, K., Schaumont, P.: 
Custom instruction support for modular defense against side-channel and fault attacks. 
In: COSADE, pp. 221--253. Springer (2020)

\bibitem{colombier2021multi}
Colombier, B., Grandamme, P., Vernay, J., Chanavat, É., Bossuet, L., de Laulani{\'e}, L., Chassagne, B.: 
Multi-spot laser fault injection setup: New possibilities for fault injection attacks. 
In: CARDIS, pp. 151--166. Springer (2021)

\bibitem{wishbone_b3}
OpenCores Organization:
WISHBONE System-on-Chip (SoC) Interconnection Architecture for Portable IP Cores, Rev.~B.3.
OpenCores (2002).
\url{https://cdn.opencores.org/downloads/wbspec\_b3.pdf}, last accessed 2025/09/04.

\bibitem{amba_axilite}
Arm Limited:
AMBA AXI4-Lite Interface (subset of AXI4).
In: AMBA AXI and ACE Protocol Specification, ARM IHI 0022E.
Arm (2013).
\url{https://documentation-service.arm.com/static/5f915b62f86e16515cdc3b1c}, last accessed 2025/09/04.

\bibitem{amba_axi}
Arm Limited:
AMBA AXI and ACE Protocol Specification, ARM IHI 0022E.
Arm (2013).
\url{https://documentation-service.arm.com/static/5f915b62f86e16515cdc3b1c}, last accessed 2025/09/04.
\end{thebibliography}
\end{document}